\documentclass[journal]{IEEEtran}
\usepackage[utf8]{inputenc} % allow utf-8 input
\usepackage[T1]{fontenc}    % use 8-bit T1 fonts
\usepackage{hyperref}       % hyperlinks
\usepackage{url}            % simple URL typesetting
\usepackage{booktabs}       % professional-quality tables
\usepackage{amsfonts, amsmath}       % blackboard math symbols
\usepackage{nicefrac}       % compact symbols for 1/2, etc.
\usepackage{microtype}      % microtypography
\usepackage{graphicx}

\ifCLASSINFOpdf
\else
\fi
\begin{document}
\title{End-to-end Networks for Supervised Single-channel Speech Separation}

\author{Shrikant~Venkataramani,~\IEEEmembership{Student Member,~IEEE,}
        Paris~Smaragdis,~\IEEEmembership{Fellow,~IEEE,}% <-this % stops a space
\thanks{S. Venkataramani  is a PhD student in the ECE department at the University of Illinois
at Urbana-Champaign e-mail: (svnktrm2@illinois.edu)}% <-this % stops a space
\thanks{P. Smaragdis holds a faculty position in Computer Science at the University of Illinois at Urbana-Champaign, and is also a senior research scientist at Adobe Research. (paris@illinois.edu)}% <-this % stops a space
\thanks{This work was supported by NSF grant 1453104.}%
\thanks{Manuscript received month day, year; revised month day, year; accepted
month day, year. Date of publication month day, year; date of current version
Month day, year.The associate editor coordinating the review of this manuscript and approving
it for publication was xxyyzz xxyyzz.}}


\markboth{IEEE Journal of Selected Topics in Signal Processing (J-STSP)}%
{Shell \MakeLowercase{\textit{et al.}}: Bare Demo of IEEEtran.cls for IEEE Journals}



\maketitle

\begin{abstract}
The performance of single channel source separation algorithms has improved greatly in recent times with the development and deployment of neural networks. However, many such networks continue to operate on the magnitude spectrogram of a mixture, and produce an estimate of source magnitude spectrograms, to perform source separation. In this paper, we interpret these steps as additional neural network layers and propose an end-to-end source separation network that allows us to estimate the separated speech waveform by operating directly on the raw waveform of the mixture. Furthermore, we also propose the use of masking based end-to-end separation networks that jointly optimize the mask and the latent representations of the mixture waveforms. These networks show a significant improvement in separation performance compared to existing architectures in our experiments. To train these end-to-end models, we investigate the use of composite cost functions that are derived from objective evaluation metrics as measured on waveforms. We present subjective listening test results that demonstrate the improvement attained by using masking based end-to-end networks and also reveal insights into the performance of these cost functions for end-to-end source separation.

\end{abstract}


% \fi
\IEEEpeerreviewmaketitle

\section{Introduction}
\IEEEPARstart{G}{iven} a mixture of multiple sources, the goal of source separation is to single-out the source of interest from the given mixture. A special case of the source separation problem is the single-channel setting wherein, the goal is to separate the source of interest given only a single mixture recording. To identify a source in the mixture, we assume that we have training data available to learn source models. We use training examples to build suitable models for the sources, and then use these models for separating unseen mixtures into its constituent sounds. \\

With the recent growing interest in deep learning and neural networks~(NN), several NN architectures have been proposed to learn models for single-channel source separation. These models are often trained discriminatively to separate a speaker of interest given a mixture of multiple speakers~\cite{isik2016single, chen2017deep, luo2018speaker, grais2017single, qian2017speech, wang2017recurrent}. Despite significant advances, these approaches predominantly rely on the use of STFTs as a front-end transformation step. The NN operates on the magnitude spectrogram of the mixture and aims to separate the speaker of interest. The separation procedure often ignores the phase component and restricts itself to magnitude spectrograms only. Thus, these approaches do not take complete advantage of all the information available in the waveforms of the signals. Considering these drawbacks, Smaragdis proposed the use of convolutional layers as a replacement to front-end and inverse STFT steps~\cite{paris_sane_2015}. Building on this, Venkataramani et.al., proposed an end-to-end source separation network that operated directly on waveforms of the mixture to learn discriminative source separation models~\cite{venkataramani_adaptive_2017}. The STFT and inverse STFT operations were replaced with convolution and transposed convolution layers. This allowed the network to operate on mixture waveforms directly and produce the waveform of the separated source at the network output. Since then, similar adaptive front-end based networks have been used to perform end-to-end separation in low-latency conditions~\cite{luo2018tasnet}. Pascual et. al., manually chunk the signal into frames to operate on the waveforms for speech enhancement~\cite{pascual2017segan}. There have also been quite a few attempts towards learning alternative end-to-end models for source separation and speech enhancement~\cite{luo2018tasnet, Fu2017Raw, rethage2018wavenet, stoller2018wave, fu2017end, qian2017speech}. However, these alternative methods experiment with bulky neural network architectures like wavenets~\cite{oord2016wavenet} or directly derive from image-processing architectures~\cite{ronneberger2015u}. Other approaches include the use of fully convolutional networks to estimate the source waveforms~\cite{Fu2017Raw, fu2017end} or use iterative phase estimation to obtain better phase estimates for the source to improve source separation performance~\cite{wang2018end}\\

With the advent of end-to-end source separation architectures, there was a surge in interest to interpret evaluation and speech intelligibility metrics as cost-functions for the networks. Fu et.al., propose a fully convolutional network trained on the Short-term Objective Intelligibility (STOI) measure~\cite{taal2010short} to perform end-to-end single channel source separation~\cite{fu2017end, fu2018end}. They demonstrated that minimizing of mean squared error (MSE) does not equate to maximizing the intelligibility of the network output. Kolbæk et.al., demonstrated that minimizing a mean squared error loss on the short-time spectral amplitudes of speech is equivalent to maximizing the STOI metric~\cite{Kolbaek2018ontheequivalence}. Despite these advances, these papers focus only on the speech intelligibility aspects of source separation. A good separation result has to be evaluated on multiple fronts: (i) Preservation of target source, (ii) Suppression of interference, (iii) Absence of additional artifacts, and (iv) Speech Intelligibility of the separated result. (i), (ii) and (iii) are measured in terms of BSS\_Eval metrics Source to Distortion ratio~(SDR), Source to Interference ratio (SIR), Source to Artifact ratio (SAR)~\cite{fevotte2005bss_eval}. Here, and in our prior work, we interpret these metrics as neural network cost-functions and propose the use of combinations of these terms to produce the best separation performance~\cite{Venkataramani2018Performance}. \\

In this paper, we build upon our prior work and propose neural network architectures for end-to-end speech separation and present three main contributions. We begin with the STFT-based source separation network and (i) generalize the architecture to develop a fully trainable end-to-end network for source separation based on adaptive front-ends. (ii) We extend this further and experiment with estimating separation masks instead of the source directly. Thus, we can now jointly optimize the estimated masks and the latent representations. And (iii) we train these end-to-end models with composite cost-functions from~\cite{Venkataramani2018Performance} that directly relate to audio quality and intelligibility. This enables us to compare the separation performance of these cost-functions over a variety of network architectures and develop insights into the nature of these cost functions using subjective listening tests.

% Maybe you don't need this...

% The remainder of the paper is organized as follows. In section~\ref{sec:ss_using_nn}, we describe the process of source separation using neural networks and develop end-to-end architectures for the same. Section~\ref{sec:cost_functions} briefly describes the performance based cost functions that can be used to train end-to-end source separation models. Next, we thoroughly evaluate the proposed models and cost functions by comparing the separation performance of these models using objective evaluations and subjective listening tests. The details and results of the experiments and subjective listening tests are covered in section~\ref{sec:experiments} and we conclude in section~\ref{sec:conclusion}. \\

\section{Source Separation using Neural Networks}
\label{sec:ss_using_nn}

In this section, we begin with the Short-time Fourier Transform based source separation network and sequentially generalize our neural network architecture to develop a fully adaptive auto-encoder based transform~(AET). Replacing the STFT by the AET allows us to develop end-to-end discriminative models for source separation, that can now operate on mixture waveforms directly, as well as allow the subsequent use of more sophisticated cost functions that are often defined in the time-domain. The presented network is initially developed as one that directly estimates the separated sources, but we also show how we can use a masking-like approach. Doing so in the AET model results in a masking operation on a learned time-frequency domain that is optimal for the goal at hand. The general nature of this approach suggests that we could possibly apply a similar strategy to develop end-to-end architectures for other audio applications.

\subsection{STFT model for Source separation}
\label{ssec:ss_using_stft}

\begin{figure}[!htb]
\centering
  \includegraphics[width=\columnwidth]{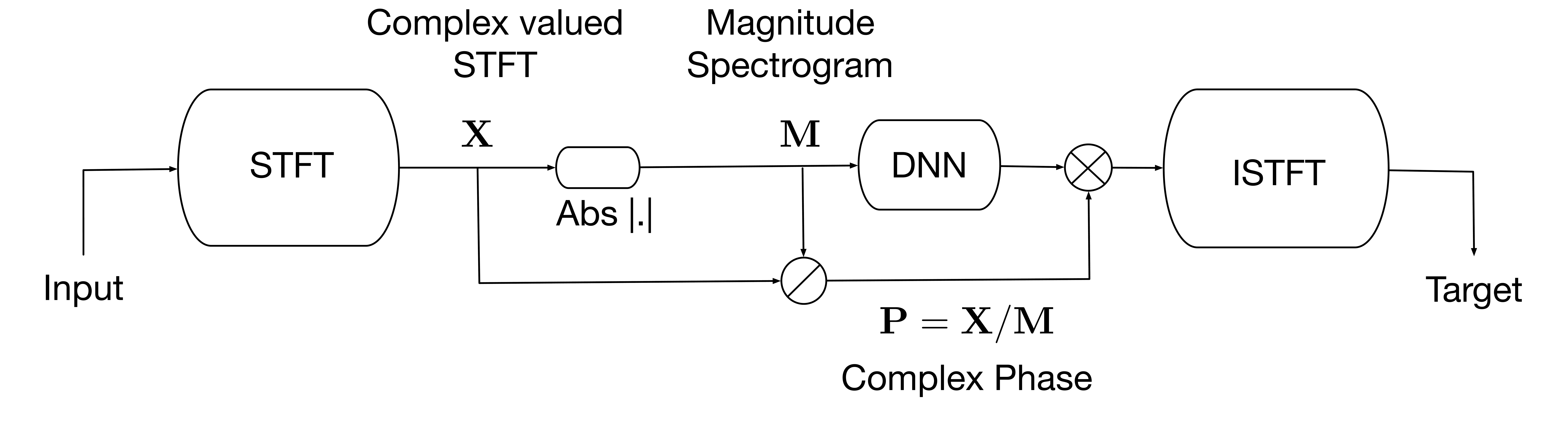}
  \caption{Source separation using the STFT. Magnitude and phase are initially separated, then a neural network operates on the magnitudes, and phase is used to modulate the network's output in order to produce the resulting source waveform.}~\label{fig:ss_stft}
\end{figure}

%% Figures
Figure~\ref{fig:ss_stft} shows the block diagram of the STFT-based source separation network. The generalized short-time transform of a time sequence~$x$ can be given as,
\begin{equation}
    \mathbf{X}(n,k) = \sum_{t=0}^{N-1} x(nh+t) \cdot w(t) \cdot b(k,t)
\label{eq:short_time_transform}
\end{equation}
In this equation, $\mathbf{X}(n,k)$ denotes the  energy of the $k^{\text{th}}$ component in the $n^{\text{th}}$ time frame of~$x$, $N$ represents the size of the lowpass window function~$w$, and $h$ represents the hop size of the short-time transform. The basis functions for the transform are denoted by $b(k,t)$. In case of STFT, $b(k,t)$ is replaced by the matrix of complex exponentials given by,
\begin{equation}
    b(k,t) = \exp\left(j \frac{2 \pi \cdot k \cdot t}{N}\right)
\label{eq:short_time_transform_basis}
\end{equation}
where integers k,t denote the frequency and time indices such that, $0 \leq k,t \leq N-1$. Thus,
\begin{align*}
    \mathbf{X}_{\text{STFT}}(n,k)  &= \sum_{t=0}^{N-1} x(nh+t) \cdot w(t) \cdot \exp\left(j \frac{2 \pi \cdot k \cdot t}{N}\right) \\
    % &= \sum_{t=0}^{N-1} x(nh+t) \cdot w(t) \cdot cos\left( \frac{2 \pi \cdot k \cdot t}{N}\right) \\
    % &+ j \cdot \sum_{t=0}^{N-1} x(nh+t) \cdot w(t) \cdot sin\left( \frac{2 \pi \cdot k \cdot t}{N}\right) \\
    % &= \text{Re}(\mathbf{X}_{\text{STFT}}(n,k)) + j \cdot \text{Im}(\mathbf{X}_{\text{STFT}}(n,k))
\end{align*}

The magnitude and phase components can be obtained as,
\begin{align*}
    \mathbf{M}_{\text{STFT}}(n,k) &= \left|\mathbf{X}_{\text{STFT}}(n,k)\right| \\
    \mathbf{P}_{\text{STFT}}(n,k) &=
    exp\left( j \cdot \angle \mathbf{X}_{\text{STFT}}(n,k) \right)\\
\end{align*}
\vspace{-2mm}

As shown in Figure~\ref{fig:ss_stft}, the magnitude component~($\mathbf{M}_{\text{STFT}}$) is now given as the input to the neural network. The network is trained to estimate the magnitude spectrogram of the source given the mixture spectrogram. The mixture phase is multiplied to the estimated source spectrogram to obtain the complex valued STFT representation of the separated source. We then transform the separated source to the time domain using the inverse STFT step.

\subsection{Smoothed STFT for source separation}
Formulating the STFT in terms of complex bases requires us to deal with complex numbers and their magnitude and phase representations. A step towards developing end-to-end models would be to interpret the STFT as a neural network operation, but complex-valued neural nets present a set of numerical challenges that we would like to avoid here. From eq.(\ref{eq:short_time_transform}), we see that $\mathbf{X}_{nk}$ is the result of a convolution operation between the current frame $[x_{nh}, x_{nh+1}, \text{\dots}, x_{nh+N-1}]$ and the $k^{th}$ basis, weighted by a window function. Thus, the real and imaginary parts of the STFT can be obtained as the output of a convolutional layer. The convolutional layer filters are set to $\mathbf{B}$ which is obtained by stacking the real and imaginary parts of the STFT bases as follows,

\[
\mathbf{B} =
\left[
\begin{array}{c}
{\cos\left( \frac{2 \pi \cdot k \cdot t}{N}\right)}_{0\leq k,t\leq N-1} \\
%\hline
{\sin\left( \frac{2 \pi \cdot k \cdot t}{N}\right)}_{0\leq k,t\leq N-1}
\end{array}
~\right]
\]

This layer's output contains the real and imaginary parts of the STFT representation, albeit not in a complex number form. Instead of subsequently computing the magnitudes, we could train the network to directly operate on these coefficients. Note that the representation now contains twice the number of coefficients as the STFT magnitudes. To obtain a representation akin to STFT magnitudes, we use an $\operatorname{abs}(\cdot)$ non-linearity at the output of the convolutional layer. \\

%% Figures
\begin{figure}[!htb]
\centering
  \includegraphics[width=\columnwidth, trim={3cm 1cm 2cm 1cm}, clip]{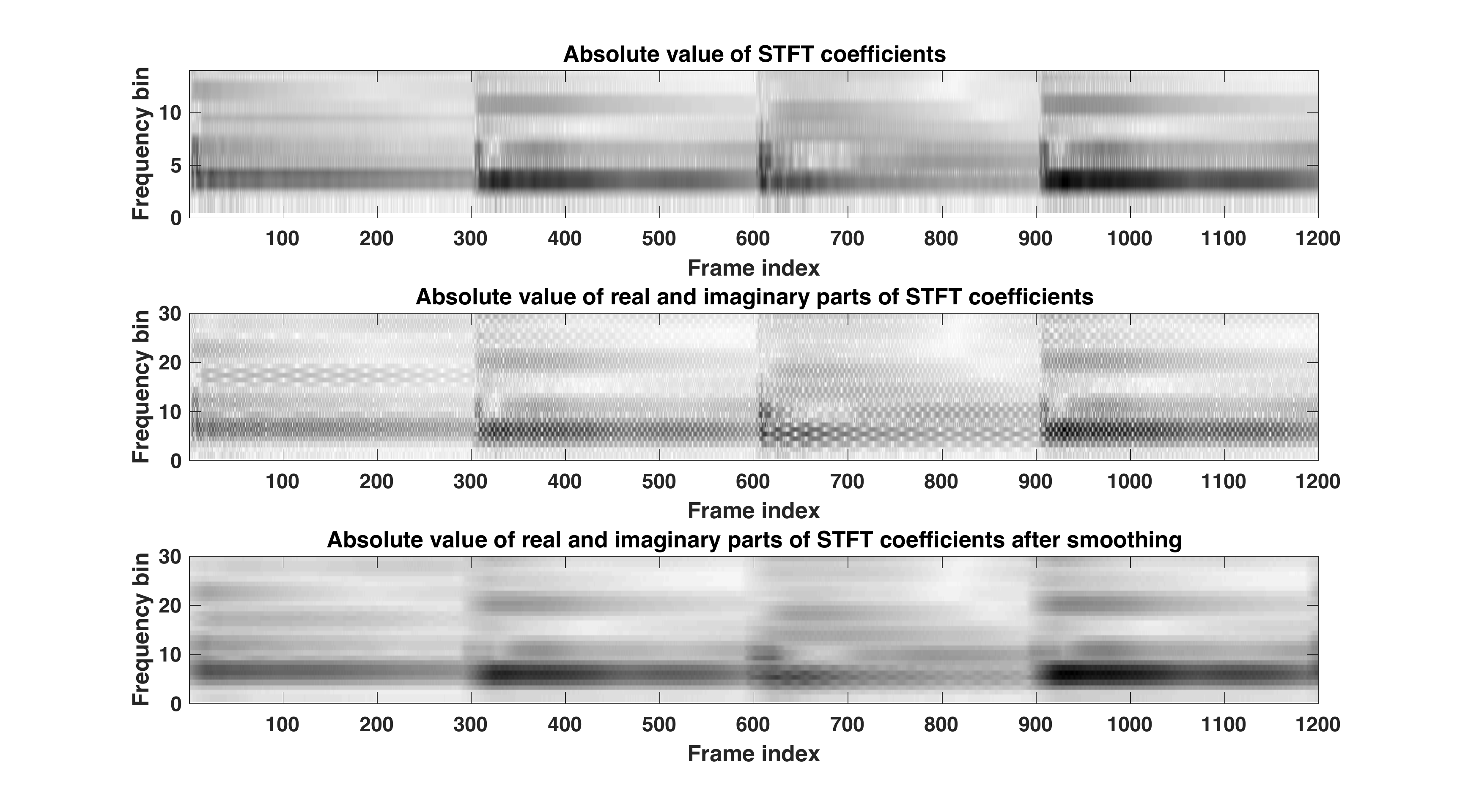}
  \caption{(a) Absolute value of the first $15$ STFT coefficients for a sequence of piano notes. (b) Modulus of equivalent real and imaginary parts of STFT coefficients. Note that we now deal with the first $30$ coefficients instead. The unsmoothed coefficients rapidly vary and have to be smoothed across time to resemble what we would expect as a magnitude spectrogram. (c) Modulus of real and imaginary parts of STFT coefficients after smoothing by a rectangular filter of length $5$.}~\label{fig:stftmags}
\end{figure}
%% Figures

Figure~\ref{fig:stftmags} compares the absolute value of the output of the front-end convolutional layer ($|\mathbf{X}|$) with the STFT magnitudes for a sequence of piano notes. We see that these coefficients vary rapidly compared to the STFT magnitudes, as a consequence of incorporating the phase information into the representation. These variations can potentially depend on the frequency of the STFT bases, the frame-size and hop-size of the front-end transform. Thus, to obtain a representation similar to the magnitude spectrogram of the STFT, we need a temporal smoothing operation on $|\mathbf{X}|$. The effect of smoothing the real and imaginary STFT coefficients using a rectangular window of duration $5$ samples is shown in Figure~\ref{fig:stftmags}. We see that the smoothed coefficients now resemble the STFT magnitudes.

The smoothing layer can be implemented as a convolutional layer that applies a convolution operation across time.
\begin{equation}
    \mathbf{M} = \left|\mathbf{X}\right|*s~
\label{eq:smoothing_stage}
\end{equation}
Here, $|\cdot|$ represents the element-wise modulus operator, $s$ represents a filter-bank of smoothing filters and $*$ denotes the one-dimensional convolution operation that is applied only across time. This smoothing operation can also be interpreted as a convolutional layer. This would allow us to learn suitable smoothing filters tailored to each frequency component. We follow the smoothing layer with a softplus non-linearity to obtain a carrier/modulator representation of the audio signal. The output of the smoothing layer is now denoted as the modulation component (M) and captures the smooth aspects of the STFT representation. The carrier component captures the information lost by the $\operatorname{abs}(\cdot)$ and smoothing operations and is given by, \begin{equation}
    \mathbf{C} = \frac{\mathbf{X}}{\mathbf{M}}
\end{equation}
where, the division is element-wise.

\begin{figure}[!htb]
\centering
  \includegraphics[width=\columnwidth]{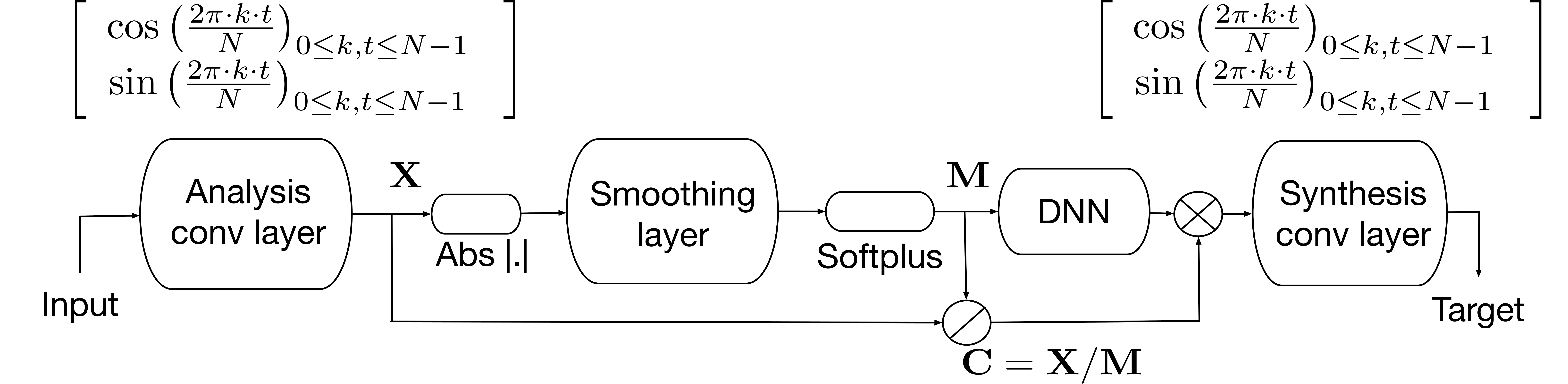}
  \caption{Taking the model in figure \ref{fig:ss_stft}, we implement it as a convolutional network with a skip connection. The analysis layer implements a real-valued version of the STFT by computing its real and imaginary components as separate dimensions. Subsequent steps extract their corresponding amplitude, process it, and recombine it with the original phase. The final layer implements the inverse transform that produces the output waveform. We additionally use a smoothing layer to compensate for the complementary modulations between the sine and cosine coefficients.}~\label{fig:ss_stft_smoothing}
\end{figure}

Figure~\ref{fig:ss_stft_smoothing} shows the block diagram of the smoothed STFT front-end for end-to-end source separation. As before, the modulation component is given as the input to the separation network. The carrier component of the mixture is multiplied with the estimated source modulation and inverted into time using a transposed convolution layer. The transposed convolutional layer is also initialized as the front-end convolutional layer in order to achieve perfect reconstruction. These layers are held fixed during training. From Figure~\ref{fig:separation_results}, we see that the smoothed STFT model is comparable to STFT magnitudes in terms of separation performance too.

%% Figures
\begin{figure*}[!htb]
\centering
  \includegraphics[width=\textwidth,trim={0.1cm 0.1cm 0.1cm 0.1cm}]{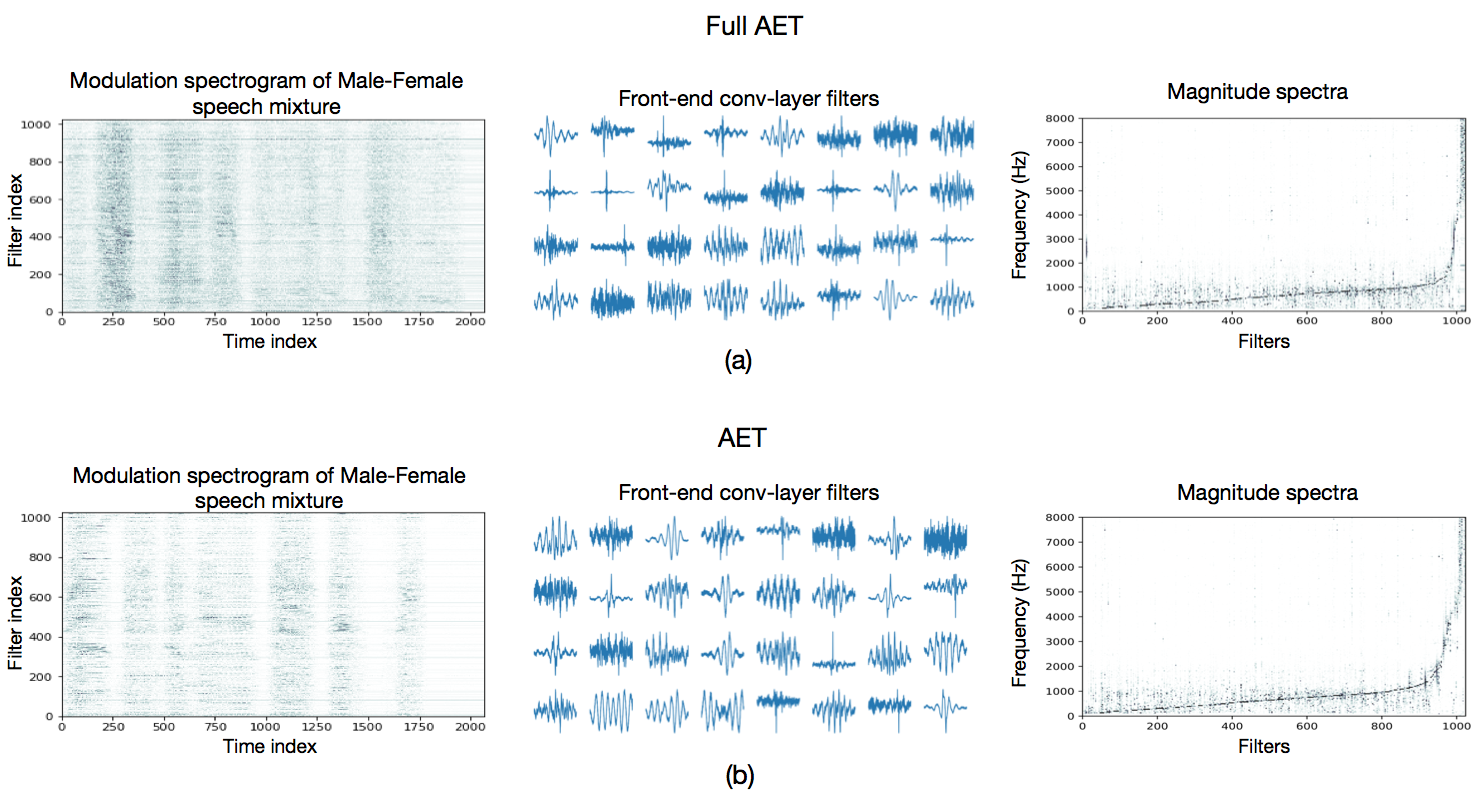}
  \caption{(a) A modulation spectrogram obtained of a speech mixture consisting of a male and female speaker, front-end convolutional layer filters, and their corresponding normalized magnitude spectra for the Full-AET model (top) (b) A modulation spectrogram obtained for a speech mixture consisting of a male and female speaker, front-end convolutional layer filters, and their corresponding normalized magnitude spectra for the AET model (bottom). The fron-end and synthesis layers share the filters in the AET model while the Full-AET model allows the front-end and synthesis layers to have independent weights. The filters are ordered according to their dominant frequency component (from low to high). In the middle subplots, we show the waveforms for a subset of the first $32$ filters.}~\label{fig:AET_bases}
\end{figure*}

\subsection{Source separation using Adaptive Front Ends}
\label{ssec:ss_using_smoothed_aet}
\begin{figure}[!htb]
\centering
  \includegraphics[width=0.95\columnwidth, trim={3cm 1.1cm 3cm 1.1cm}, clip]{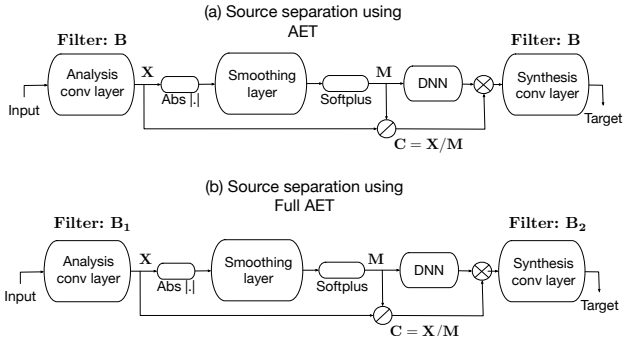}
  \caption{Block diagrams of end-to-end source separation networks using the AET (a), and the Full-AET (b). In contrast to before, these models have a trainable front-end as opposed to using a fixed Fourier-related filterbank. In the case of the Full-AET, the final layer has an independent trainable filterbank, whereas the AET is using the transpose of the learned analysis filterbank.}~\label{fig:ss_aet}
\end{figure}

Having developed an end-to-end separation network that operates on mixture waveforms, we can now make the convolutional and smoothing layers learnable to obtain an end-to-end fully adaptive neural network. Making the front-end and smoothing layers adaptive allows the network to learn basis and smoothing functions directly from the raw waveform of the signal. Thus, these functions would be optimal for the separation task. The transposed convolutional layer acts as an adaptive reconstruction step that transforms the audio signal back into the waveform domain. We refer to this as the auto-encoder transform (AET) front end. This allows us the possibility to explore two possible configurations of the AET networks. In the first configuration, the front-end and synthesis convolutional layers share the same filters. The second configuration is to allow the front-end and synthesis convolutional layers to be independently trainable. In the remainder of the paper, we refer to these configurations as the AET and Full-AET respectively. Figure~\ref{fig:ss_aet} gives the block diagrams of the AET and Full-AET architectures.
\\

\vspace{-1mm}

Figure~\ref{fig:AET_bases} shows the AET and Full-AET front-end filters learned on Male-Female mixtures. We plot these filters in time and frequency. We see that the filters are frequency selective like the STFT. Also, the filters are concentrated towards the lower frequencies and spread-out at the higher frequencies, similar to the Mel-filter bank. Unlike the STFT, the filters of the front-end and synthesis convolutional layers are not restricted to be orthogonal or inverse versions of each other. In addition, the stride of the convolutional layers also plays a role in developing trainable architectures. The effect of striding on AET and Full-AET models is discussed in section~\ref{sssec:effect_of_stride} and Figure~\ref{fig:SDRvsStride}.

\subsection{Source separation using Mask Estimation}
\label{ssec:ss_using_masking}
\begin{figure}[!htb]
\centering
  \includegraphics[width=\columnwidth]{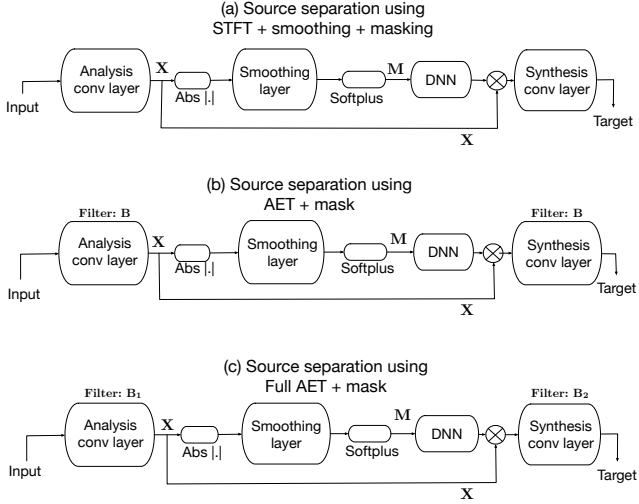}
  \caption{Masking based architectures for end-to-end speech separation. These networks are derived from the STFT, AET and Full-AET architectures by modifying their skipped connections to resemble masking. This allows the AET and Full-AET versions to learn the latent representations and source masks jointly.}~\label{fig:ss_mask}
\end{figure}

As seen previously, the carrier component $\mathbf{C} = \frac{\mathbf{X}}{\mathbf{M}}$ retains the information lost by the $\operatorname{abs}(\cdot)$ .and smoothing operations. The carrier component is multiplied to the output of the separation network, along the lines of the STFT phase. This can be interpreted as a skipped connection that benefits the transport of gradients through the network~\cite{he2016deep}. A popular source separation strategy is to estimate the ideal-ratio mask (IRM) for the source of interest~\cite{hummersone2014ideal}. The IRM contains a value between $0$ and $1$ for each time-frequency bin and is applied as an element-wise multiplication on the complex-valued STFT of the mixture to estimate the complex-valued STFT of the source. In the case of neural networks, the estimation of IRMs has been shown to be beneficial for source separation~\cite{huang2015joint}. We can simplify the nature of our skipped connections by adopting a similar mask estimation strategy. The mask estimation equivalents of the proposed architectures are shown in Figure~\ref{fig:ss_mask}. In these figures, the DNN is assumed to have a sigmoid non-linearity in its last layer, and generates a value between $0$ and $1$ at its output. Note that, in Figure~\ref{fig:ss_mask} (b) and (c), the masks are applied on the adaptive latent representation of the signal. We also note that these networks are still trained in an end-to-end manner similar to the above architectures. \\

\vspace{-1mm}

%% Figures
% \begin{figure*}[!htb]
% \centering
%   \includegraphics[width=\textwidth, trim={0.1cm 0.1cm 0.1cm 0.1cm}, clip]{Figs/BlockDiagrams.pdf}
%   \caption{Block diagrams of the proposed architectures. (a) STFT model (b) STFT smoothed (c) STFT smoothed mask (d) AET (e) AET mask (f) Full AET (g) Full AET mask.}~\label{fig:block_diagrams}
% \end{figure*}

Combining mask estimation with the proposed networks, we now have the following $7$ architectures for source separation. We list the architectures and also briefly point out the salient aspects of each model. The corresponding block diagram figures are also linked in the following brackets. \\

(i) STFT: uses STFT magnitudes for source separation~(Figure~\ref{fig:ss_stft}). \\

(ii) STFT smoothed: uses smoothed real and imaginary STFT coefficients for source separation~(Figure~\ref{fig:ss_stft_smoothing}).  \\

(iii) STFT smoothed mask: uses smoothed real and imaginary STFT coefficients and estimates masks for source separation~(Figure~\ref{fig:ss_mask}(a)).\\

(iv) AET: uses auto-encoder like convolutional layers to learn adaptive bases. The front-end and synthesis layers share the weights~(Figure~\ref{fig:ss_aet}(a)). \\

(v) AET mask: uses auto-encoder like convolutional layers to learn adaptive bases. The front-end and synthesis layers share the weights. Estimates a mask at the output of the separation network.~(Figure~\ref{fig:ss_mask}(b)) \\

(vi) Full-AET: uses auto-encoder like convolutional layers to learn adaptive bases. The front-end and synthesis layers have independent weights~(Figure~\ref{fig:ss_aet}(b)). \\

(vii) Full-AET mask: uses auto-encoder like convolutional layers to learn adaptive bases. The front-end and synthesis layers have independent weights. Estimates a mask at the output of the separation network~(Figure~\ref{fig:ss_mask}(c)). \\

% The block diagrams for these models are shown in Figure~\ref{fig:block_diagrams}.

In the next section, we look into appropriate cost functions for training these neural networks directly on the raw waveforms.

\section{Cost-functions for end-to-end speech separation}
\label{sec:cost_functions}
Previously, magnitude spectrograms have been interpreted as probability density functions of random variables with varying characteristics. This has led to the use of divergence based cost-functions for single channel source separation. Some of these examples include~Mean squared error (MSE)~\cite{wang2018alternative}, KL Divergence~\cite{smaragdis2017neural, nugraha2016multichannel} and IS divergence~\cite{nugraha2016multichannel}. However these are often used as proxies to the performance metrics we ultimately measure, which are almost always waveform-based. For end-to-end architectures, statistical metrics like mean squared error~\cite{stoller2018wave, venkataramani_adaptive_2017} and l-1 loss~\cite{wang2018end, rethage2018wavenet} have been tried. Cross-entropy and its modified versions have also been tried out~\cite{pascual2017segan}. In source separation however, we most often evaluate the performance of source separation algorithms using BSS\_Eval metrics viz., SDR, SIR, SAR~\cite{fevotte2005bss_eval} and Short-term objective Intelligibility (STOI)\cite{taal2010short} metrics. Thus, a logical step is to interpret these metrics as cost-functions themselves.

In~\cite{venkataramani_adaptive_2017}, we proposed the use of SDR as a cost-function that can be applied to waveforms of end-to-end systems. We showed that the use of SDR resulted in a significant improvement in separation performance. SDR has also been used as a cost-function for end-to-end source separation in~\cite{luo2018tasnet}. In~\cite{Venkataramani2018Performance}, we expanded upon the premise to interpret SIR, SAR and STOI as cost-functions for end-to-end separation. This allowed us to experiment with combinations of these to improve separation performance for the Full-AET architecture. In this paper, we generalize the study and its conclusions by comparing these cost-functions across architectures. First, we give a brief summary on using BSS\_Eval and STOI as cost functions. In these cost functions, we  denote  the network output waveform as $\mathbf{x}$. This output should ideally match the source waveform $\mathbf{y}$ and suppress the interference~$\mathbf{z}$. Thus, $\mathbf{y}$ and $\mathbf{z}$ are fixed constants with respect to the optimization. The detailed derivations and descriptions can be found in~\cite{Venkataramani2018Performance}.

\subsection{BSS\_Eval Cost Functions}
The distortions in the network output $\mathbf{x}$ can result from the effects of the interfering source $\mathbf{z}$ or from the artifacts introduced by the processing algorithm. SDR incorporates the overall effect of distortions in the network output. The effects of the interfering sources are observed in the SIR score. The processing artifacts are measured by SAR. Maximizing SDR with respect to $\mathbf{x}$ maximizes the correlation between $\mathbf{x}$ and $\mathbf{y}$ and simultaneously minimizes the energy of $\mathbf{x}$. This can be given as,

\begin{align*}
     \max \operatorname{SDR}(\mathbf{x}, \mathbf{y})
     & \propto \min \frac{\langle \mathbf{xx} \rangle}{\langle \mathbf{xy} \rangle^2}
\end{align*}

Maximizing the SIR maximizes the correlation of the network output $\mathbf{x}$ with the desired source $\mathbf{y}$ and minimizes the correlation between $\mathbf{x}$ and the interference $\mathbf{z}$.

\begin{align*}
    \max \operatorname{SIR}(\mathbf{x}, \mathbf{y}, \mathbf{z})
    \propto \min \frac{\langle \mathbf{xz} \rangle^2}{ \langle \mathbf{xy}\rangle^2}
\end{align*}

Removing the minimum energy constraint makes the network focus predominantly on time-frequency bins that are dominated by $\mathbf{y}$ and do not contain $\mathbf{z}$. Thus, SIR as a cost-function needs to be supported by SAR given as,

\begin{align*}
     \max \operatorname{SAR}(\mathbf{x}, \mathbf{y}, \mathbf{z})
     & \propto \min \frac{\langle \mathbf{xx} \rangle}
     {\frac{\langle \mathbf{xy} \rangle^2}{\langle \mathbf{yy} \rangle} + \frac{\langle \mathbf{xz} \rangle^2}{\langle \mathbf{zz} \rangle}}
 \end{align*}

 The SAR cost function does not distinguish between $\mathbf{y}$ and $\mathbf{z}$ and cannot be used to train a network to separate the source from the interferences. However, in combination with SIR, it can be used for end-to-end separation.

\subsection{Short Term Objective Intelligibility (STOI)}
The drawback of BSS\_Eval metrics is that, they do not necessarily reflect the amount of intelligibility of the resulting output. STOI is a popular metric that correlates with subjective speech intelligibility. The first step of computing the STOI metric is to transform the audio signals into the time-frequency domain using a $512$-point STFT with a ``Hann'' window size of $256$ samples and a hop of $128$ samples. The STFT representations are transformed into an octave band representation using $15$ (1/3)rd octave bands that extend upto $10000 Hz$. These steps are applied on the network output $\mathbf{x}$ and the source of interest $\mathbf{y}$ to obtain $\mathbf{\hat{X}}$ and $\mathbf{\hat{Y}}$ respectively. Here, $\mathbf{\hat{X}}_{j, m}$ corresponds to the energy of $\mathbf{x}$ in the $j$~-th one-third octave band at the $m$~-th time frame.

Given $\mathbf{\hat{X}}$ and $\mathbf{\hat{Y}}$, the intermediate STOI measure for one bin, denoted by $d_{j,m}$ depends on a neighborhood of $N$ previous bins. To do so, we construct new vectors $\mathbf{X}_{j, m}$ and $\mathbf{Y}_{j, m}$ consisting of $N = 30$ previous frames before the $m$~-th time frame as follows,
\begin{align*}
     \mathbf{X}_{j, m} = [\mathbf{\hat{X}}_{j, m - N + 1}, \mathbf{\hat{X}}_{j, m - N + 2}, ..., \mathbf{\hat{X}}_{j, m}]^T
\end{align*}

We scale and clip $\mathbf{X}_{j,m}$ to construct $\mathbf{\bar{X}}_{j,m}$ as follows,
\begin{align*}
     \mathbf{\bar{X}}_{j, m}(n) = \min \left (\frac{\lVert \mathbf{Y}_{j, m} \rVert}{\lVert \mathbf{X}_{j, m} \rVert} \mathbf{X}_{j, m}(n), (1 + 10^{-\beta/20}) \mathbf{Y}_{j, m}(n) \right)
\end{align*}

In this equation, $\mathbf{X}_{j, m}(n)$ denotes the $n$~-th value of $\mathbf{X}_{j, m}$ and $\beta$ is set to $-15$~dB.
The intermediate intelligibility matrix, $d_{j, m}$ can be calculated as the correlation between $\mathbf{\bar{X}}_{j,m}$ and $\mathbf{Y}_{j,m}$. This can be written as,
 \begin{align*}
     d_{j, m} = \frac{(\mathbf{\bar{X}}_{j, m} - \mathbf{\mu_{\bar{X}}}_{j, m})^T(\mathbf{Y}_{j, m} - \mathbf{\mu_{Y}}_{j, m})}
     {\lVert \mathbf{\bar{X}}_{j, m} - \mathbf{\mu_{\bar{X}}}_{j, m} \rVert \cdot \lVert \mathbf{Y}_{j, m} - \mathbf{\mu_{Y}}_{j, m} \rVert}
 \end{align*}
 The overall STOI metric is given as the average of $d_{j,m}$ over all time-frames and all octave bands. Thus, maximizing STOI as a cost-function maximizes the correlation between the octave band representations of $\mathbf{x}$ and $\mathbf{y}$.\\

 Based on our previous work~\cite{Venkataramani2018Performance}, we can draw the following conclusions based on the use of these cost functions. Each function captures different salient aspects of source separation and a good strategy to improve separation performance is to use a combination of cost functions. For the Full-AET separation network, $0.75\times\operatorname{SDR} + 0.25\times\operatorname{STOI}$ gives the best performance overall in terms of preserving the target source and intelligibility of the separated result. Also, $0.5\times\operatorname{SIR} + 0.5\times\operatorname{SAR}$ gives the best results in suppressing the interference among the cost-functions tested. In this paper, we compare the use of $\operatorname{MSE}$, $\operatorname{SDR}$, $0.75\times\operatorname{SDR} + 0.25\times\operatorname{STOI}$ and $0.5\times\operatorname{SIR} + 0.5\times\operatorname{SAR}$ across multiple architectures in order to generalize these conclusions. The details of the listening tests and results are given in section~\ref{subsec:separation_experiments}.

\section{Experiments}
\label{sec:experiments}
Having developed a strategy to learn end-to-end source separation models, we turn our attention to evaluating the proposed models and cost functions. We divide the experiments into two segments: (i) Evaluating end-to-end AET networks (ii) Evaluating cost functions.

\subsection{Separation Experiments}
\label{ssec:separation_experiments}
In this segment, we describe a series of experiments to compare the separation performances of the proposed end-to-end source separation architectures. We first describe the dataset for these experiments. \\

\subsubsection{Dataset}
\label{sssec:dataset}
For our experiments, we used the training-folders (si\_tr\_s) of the Wallstreet Journal 0 (WSJ0)~\cite{wsj} corpus. $25$ male and $25$ female speakers were selected at random for training our networks. For the evaluations, we selected a different set of $10$ male and $10$ female speakers. Thus, the evaluation is truly speaker independent because, the evaluation speakers and mixtures were not a part of the training set. For training, we randomly selected a pair of male and female speakers from the training set speakers. An audio utterance was selected at random for each speaker and a $2$-sec snippet was drawn at random from each utterance. These snippets were mixed at $0$~dB to generate the mixture waveform. The female utterance was selected as the source of interest and all the networks were trained to separate the female speech from the mixture. For the evaluations, we constructed similar male-female speaker mixtures at $0$~dB from speakers selected for testing.\\

\subsubsection{Experimental Setup}
\label{sssec:experimental_setup}
We now describe the parameters of our network architecture. For the STFT front-end, the coefficients were computed using a Hann window of duration $1024$~samples at a hop of $16$~samples. For the smoothed STFT version, the front-end consisted of $1024$~coefficients with $512$~real and $512$~imaginary components each. For these coefficients, a smoothing operation of duration $5$~samples resulted in a smooth modulation spectrogram as shown in Figure~\ref{fig:stftmags}. Thus, smoothing layer filters were set to a length of $5$ so as to average over $5$ previous frames. For the AET networks, a front-end convolutional layer was set to have a stride of $16$~samples so as to enable a fair comparison with the fixed front-end networks. The separation network consisted of a cascade of $3$ dense layers, each followed by a softplus non-linearity. For the masking based networks, a sigmoid non-linearity was used in the last dense layer. Each of these hidden layers were selected to have a depth of $512$ neurons each. For these experiments, all the networks were trained using SDR as the cost-function. \\

%% Figures
\begin{figure*}[!htb]
\centering
  \includegraphics[width=0.95\textwidth, trim={16cm 7cm 10cm 2cm}, clip]{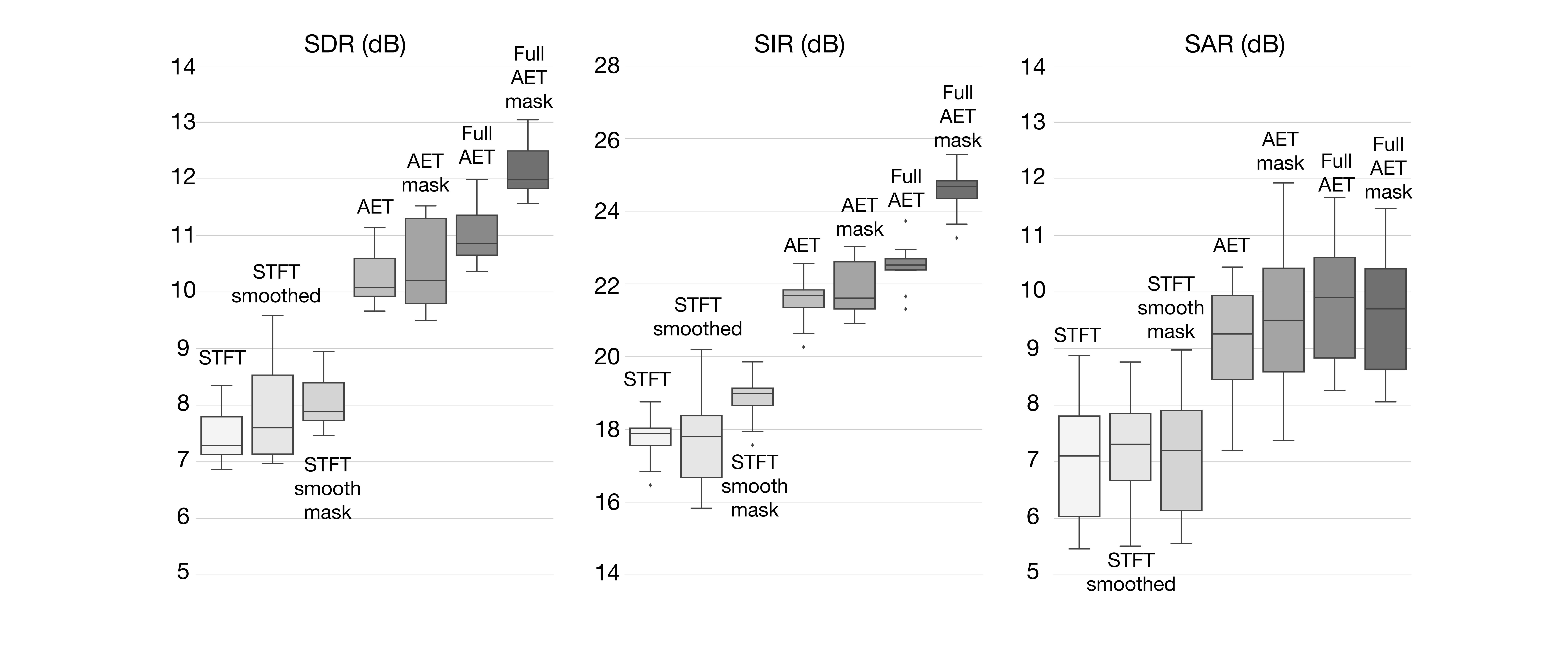}
  \caption{Comparison of the separation performance of the proposed models in terms of BSS\_Eval metrics: SDR, SIR and SAR. We see that estimating a mask instead of directly estimating the source improves separation performance. The use of end-to-end neural networks to learn adaptive front-ends also boosts separation performance.}~\label{fig:separation_results}
\end{figure*}

\subsubsection{Source separation performance}
\label{sssec:Source_separation_performance}
The first experiment aims to compare the separation performance of the proposed end-to-end architectures. We construct a training dataset as outlined in section~\ref{sssec:dataset} of $200$ minute duration. All the networks were trained on this dataset and evaluated on a testset of duration $20$ minutes. We compare the models in terms of the BSS\_Eval metrics: SDR, SIR and SAR computed on the test set. Figure~\ref{fig:separation_results} gives the results of the experiment in the form of a box plot that shows the median (solid line in the middle) and the $25^{th}$ percentile and the $75^{th}$ percentile points as the box extremities. \\

We observe that the smoothed STFT model is comparable to the STFT model in terms of median separation performance. However, it results in a higher variance compared to the STFT front-end. Another consistent feature of these experiments is that, estimating a separation mask results in a significant improvement compared to directly estimating the sources. This can be seen from the following pairs of violin-plots: STFT smoothed \& STFT smoothed mask, AET \& AET mask, Full-AET \& Full-AET mask. The results also demonstrate a consistent improvement in the separation performance as we move towards a completely trainable network with adaptive front-ends, with the Full-AET mask network significantly outperforming the rest.  \\

\begin{figure*}[ht!]
\centering
  \includegraphics[width=0.9\textwidth, trim={4cm 0.1cm 4.3cm 0.1cm}, clip]{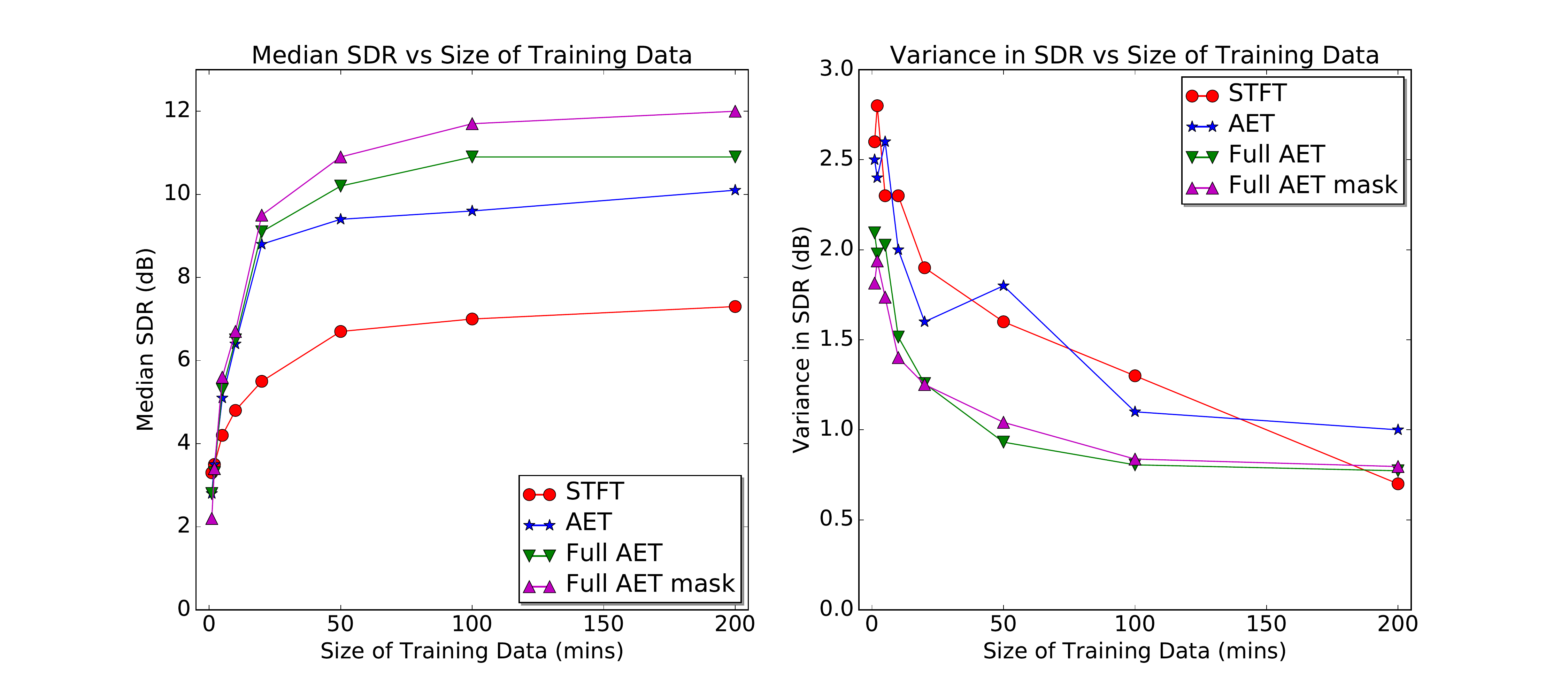}
  \caption{(Left) A plot of separation performance in terms of median SDR value at varying amounts of training data. (Right) A plot of variance of SDR at varying amount of training data. The variance of the separation results decreases with increasing training data size for all the models. The improvement in separation performance is not very pronounced beyond $100$ minutes of training data. Here again, the Full-AET mask model outperforms the rest of the models.}~\label{fig:BigDataExperiment_AdaptiveFrontEnds}
\end{figure*}

\subsubsection{Effect of training data size}

The performance of neural network based approaches are affected by the amount of data available for training. The next experiment aims to quantify the separation performance of end-to-end source separation at varying amounts of training data. We construct training sets of duration $1, 2, 5, 10, 20, 50, 100, 200$ minutes and a test set of duration $20$ minutes. The generation of these datasets has been described in section~\ref{sssec:dataset}. For simplicity and ease of understanding, we show the results of the experiment for a few selected representative networks viz., STFT, AET, Full-AET and Full-AET mask. Since SDR gives the overall metric for measuring separation performance, we plot the comparison in terms of SDR. Figure~\ref{fig:BigDataExperiment_AdaptiveFrontEnds} shows the change in performance of the networks with an increase in training data size. The plots show the median SDR value (left) and the variance of SDR (right) at varying amounts of training data. \\

We see that increasing the training data size improves the separation performance consistently for all the models. Beyond $100$ mins of training data, the improvement in separation performance is not very prominent for all the models. As expected, higher amount of training data favor end-to-end architectures as they show significantly better separation on speech mixtures. \\

\subsubsection{Effect of stride}
\label{sssec:effect_of_stride}
One of the advantages of the STFT is that STFTs allow us to perfectly reconstruct the audio signal from a highly sub-sampled time-frequency representation. Perfect reconstruction is achieved by using overlapping windows and specific values of hop~\cite{allen1977unified}. In case of the adaptive front-ends the STFT hop has been replaced by the stride of the front-end convolutional layers. We now consider the effect of strides on source separation performance. As before, we plot the variation in median SDR for different values of stride for  STFT, AET, Full-AET and Full-AET mask models. To have a fair comparison with the STFT model, we show the plots for hop/stride values that give perfect reconstruction for a window-size of $1024$ samples. The advantage of using strided convolutional layers is that it reduces the number of computations required. \\

\begin{figure}[!htb]
\centering
  \includegraphics[width=\columnwidth]{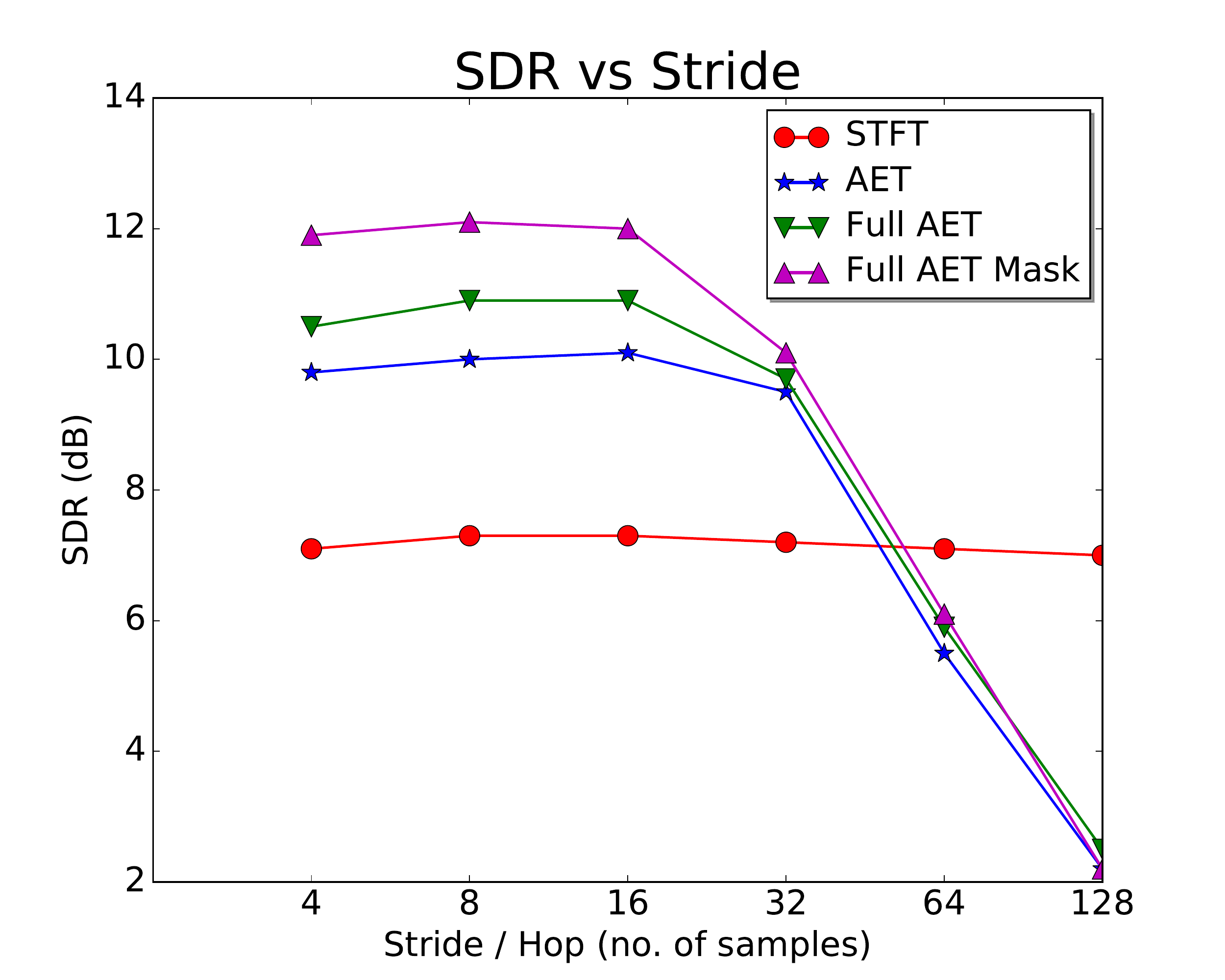}
  \caption{Plot of separation performance vs convolutional layer stride/hop. In case of the STFT, the separation performance remains consistent as it continues to have perfect reconstruction. In case of these end-to-end models, the separation performance rolls off significantly if we increase the stride beyond $32$ samples. Thus, the adaptive latent representation cannot be as highly subsampled as the STFT.}~\label{fig:SDRvsStride}
\end{figure}
\vspace{-1mm}

Figure~\ref{fig:SDRvsStride} shows the variation in SDR for varying values of stride. There are a few key differences in strided convolutions as compared to STFT front-ends. (i) STFTs allow us to perfectly reconstruct the audio signal in time only for specific values of the hop size~\cite{allen1977unified}. In the case of end-to-end networks, strided convolutional layers can be trained to reconstruct the waveforms for all values of stride upto $32$ samples. The window shape and bases are adaptively learned to suit the stride without restrictions. (ii) There is no significant improvement or deterioration in separation performance for strides upto $16$ samples. Increasing the stride beyond $32$ samples results in a sharp fall in separation performance. However, the STFT continues to perfectly reconstruct the audio signal if the hop size is appropriately chosen. \\

\begin{figure*}[!htb]
\centering
  \includegraphics[width=0.94\textwidth, trim={0cm 0.1cm 0cm 0.3cm}, clip]{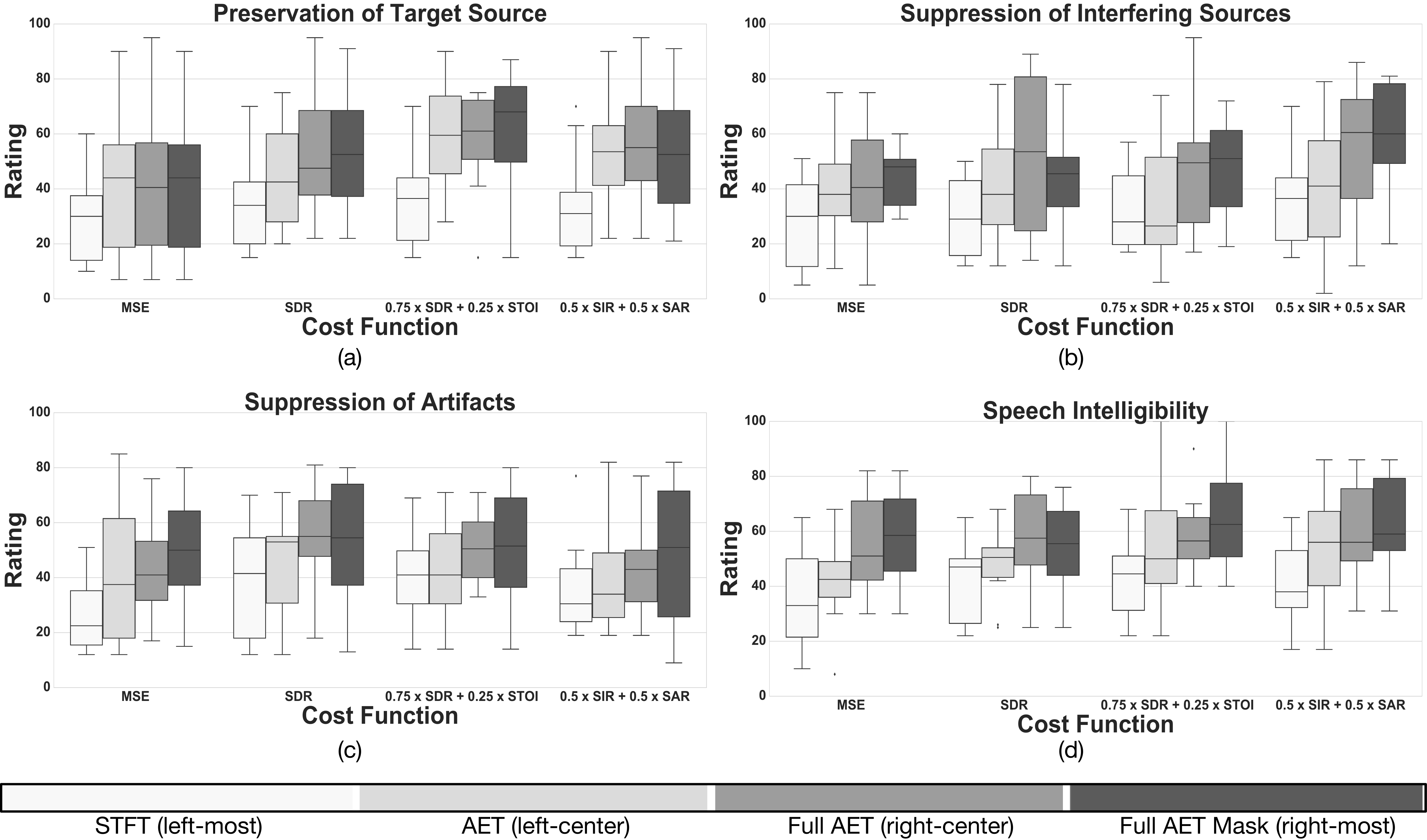}
  \caption{Subjective listening test scores for tasks (a) Preservation of target source. (b) Suppression of interfering sources. (c) Suppression of artifacts (d) Speech intelligibility for the following models: STFT (left bar), AET (left-center bar), Full-AET (right-center bar) and Full-AET mask (right bar). The horizontal axis shows the cost function used for training. The distribution of scores is presented as a box-plot where, the solid line in the middle give the median value and the extremities of the box give the $25^{\text{th}}$ and $75^{\text{th}}$ percentile values.}~\label{fig:Listening_test}
\end{figure*}
\vspace{-1mm}
%% Figures

\subsection{Subjective Listening Tests}
\label{subsec:separation_experiments}
For the second segment of our experiments, the goal is to compare and rate the use of different cost functions for end-to-end source separation. As before, we restrict our attention to the following $4$ models: STFT, AET, Full-AET and Full-AET mask.  The networks were trained on a training set of duration $200$ minutes using different cost functions. The experimental setup used is described in section~\ref{sssec:experimental_setup}. Since the cost functions use BSS\_Eval and STOI metrics themselves, a fair assessment would require the use of subjective listening tests to evaluate the performance of these cost functions. In these tests, we compare the separation performance under the following costs: \\
(i) Mean squared error \\
(ii) $\operatorname{SDR}$ \\
(iii) $0.75\times\operatorname{SDR} + 0.25\times\operatorname{STOI}$ \\
(iv) $0.75\times\operatorname{SIR} + 0.25\times\operatorname{SAR}$ \\

(iii) and (iv) were selected based on the fact that they gave the best separation results in subjective tests conducted in our earlier work in~\cite{Venkataramani2018Performance}. We scale the value of each cost-function to unity before starting training. This was done to appropriately implement the weighting of terms in case of composite cost-functions. We used the CAQE toolkit~\cite{cartwright_fast_2016} to conduct our subjective listening tests by recruiting listeners over Amazon Mechanical Turk (AMT). Subjective listening tests conducted over CAQE have been shown to yield results consistent with those performed under controlled environments.\\

\subsubsection{Recruiting Participants}
We now briefly describe our approach to recruiting participants over AMT. Listeners that were atleast $18$ years of age and with no previous history of hearing impairments were recruited for the listening tests. The listeners were tested with a few brief hearing tests where the listener had to correctly identify the number of sinusoidal tones in an audio played to them. The pariticipants were afforded two attempts to get the answers correct, otherwise their response was rejected. We recruited a total of $180$ participants over AMT for these tests. \\

\subsubsection{Experiments}
The cost functions and networks were compared in terms of four distinct separation tasks: (i) Preservation of target source, (ii) Suppression of interference, (iii) Absence of additional artifacts, and (iv) Speech intelligibility. The objective equivalent metrics for these tasks are SDR, SIR, SAR and STOI respectively. In addition to a description of the task, the listeners were trained by providing them with an example mixture, the clean sources and $3$ good and bad separation examples for each task. To identify these good and bad examples, the equivalent objective metric scores were used and audio examples with exceptionally high or low scores were selected at random. Listeners were asked to provide a mean opinion score between $0-100$ for each task. After the training step, listeners were asked to rate $16$ audio examples, one for each cost function and model pair. This was done to have a level of consistency across these scores for every set. The listeners were also provided with the clean source to be used as the baseline. The audio examples were unlabelled and randomized to avoid biasing the listener scores. \\

\subsubsection{Results and Discussion}
Figure~\ref{fig:Listening_test} gives the mean-opinion scores of the subjective listening tests as a bar-plot. The median score (solid line at the center) and the $25^{th}$, $75^{th}$ percentile values (box edges) are shown in the plots. From these plots, we see that moving towards an end-to-end separation networks and masking based separation networks improves separation performance. This is observed in the fact that the best overall results are obtained by the Full-AET masking model. The subjective scores corroborate the objective evaluations performed in section~\ref{sssec:Source_separation_performance}. These listening tests also allow us to generalize the claims of our preliminary listening test performed in~\cite{Venkataramani2018Performance}. We observe that using a combination of SDR and STOI gives better separation performance in terms of preserving the target source and speech intelligibility. SDR continues to be cost function of choice when it comes to minimal artifacts for source separation across all the four models. Similarly, a combination of SIR and SAR as the cost function gives the best suppression of the interfering sources.

\section{Conclusion}
\label{sec:conclusion}
In this paper, we have shown how we can generalize existing STFT based source separation networks to develop end-to-end neural networks for source separation. We extend this further to propose mask based end-to-end networks that allow us to simultaneously optimize over latent representations of the waveform and the estimation masks. Finally, we show how these architectures allow us to utilize cost functions that directly map to perceived performance. Our experiments reveal that these models outperform existing architectures significantly in terms of separation performance. The results of listening tests corroborate to the superiority of masking based end-to-end networks over existing architectures. Through the listening tests, we also compared the separation results of performance based cost functions for end-to-end separation and draw the following conclusions: (i) A SDR-STOI combination gives the best separation performance in terms of preserving the target source. (ii) Maximizing the SDR also produces the least artifacts among the cost-functions tested. (iii) A combination of SIR and SAR gives the best suppression of interfering sources.

% or


% \appendices

% % you can choose not to have a title for an appendix
% % if you want by leaving the argument blank
% \section{}

% \section*{Acknowledgment}
% This work was supported by NSF grant 1453104.

\ifCLASSOPTIONcaptionsoff
  \newpage
\fi



%\bibliographystyle{IEEEtran}

\bibliographystyle{IEEEtran}
\bibliography{refs}

\begin{thebibliography}{10}
\providecommand{\url}[1]{#1}
\csname url@samestyle\endcsname
\providecommand{\newblock}{\relax}
\providecommand{\bibinfo}[2]{#2}
\providecommand{\BIBentrySTDinterwordspacing}{\spaceskip=0pt\relax}
\providecommand{\BIBentryALTinterwordstretchfactor}{4}
\providecommand{\BIBentryALTinterwordspacing}{\spaceskip=\fontdimen2\font plus
\BIBentryALTinterwordstretchfactor\fontdimen3\font minus
  \fontdimen4\font\relax}
\providecommand{\BIBforeignlanguage}[2]{{%
\expandafter\ifx\csname l@#1\endcsname\relax
\typeout{** WARNING: IEEEtran.bst: No hyphenation pattern has been}%
\typeout{** loaded for the language `#1'. Using the pattern for}%
\typeout{** the default language instead.}%
\else
\language=\csname l@#1\endcsname
\fi
#2}}
\providecommand{\BIBdecl}{\relax}
\BIBdecl

\bibitem{isik2016single}
Y.~Isik, J.~L. Roux, Z.~Chen, S.~Watanabe, and J.~R. Hershey, ``Single-channel
  multi-speaker separation using deep clustering,'' 2016.

\bibitem{chen2017deep}
Z.~Chen, Y.~Luo, and N.~Mesgarani, ``Deep attractor network for
  single-microphone speaker separation,'' in \emph{2017 IEEE International
  Conference on Acoustics, Speech and Signal Processing (ICASSP)}, March 2017,
  pp. 246--250.

\bibitem{luo2018speaker}
Y.~Luo, Z.~Chen, and N.~Mesgarani, ``Speaker-independent speech separation with
  deep attractor network,'' \emph{IEEE/ACM Transactions on Audio, Speech, and
  Language Processing}, vol.~26, no.~4, pp. 787--796, April 2018.

\bibitem{grais2017single}
E.~M. Grais and M.~D. Plumbley, ``Single channel audio source separation using
  convolutional denoising autoencoders,'' \emph{arXiv preprint
  arXiv:1703.08019}, 2017.

\bibitem{qian2017speech}
K.~Qian, Y.~Zhang, S.~Chang, X.~Yang, D.~Flor{\^e}ncio, and
  M.~Hasegawa-Johnson, ``Speech enhancement using bayesian wavenet,'' 2017.

\bibitem{wang2017recurrent}
Z.-Q. Wang and D.~Wang, ``Recurrent deep stacking networks for supervised
  speech separation,'' in \emph{Acoustics, Speech and Signal Processing
  (ICASSP), 2017 IEEE International Conference on}.\hskip 1em plus 0.5em minus
  0.4em\relax IEEE, 2017, pp. 71--75.

\bibitem{paris_sane_2015}
\BIBentryALTinterwordspacing
P.~Smaragdis, ``Nmf? neural nets? it’s all the same.'' 2015, speech and Audio
  in the Northeast (SANE). [Online]. Available:
  \url{http://www.merl.com/events/sane2015}
\BIBentrySTDinterwordspacing

\bibitem{venkataramani_adaptive_2017}
S.~Venkataramani, J.~Casebeer, and P.~Smaragdis, ``Adaptive front-ends for
  end-to-end source separation,'' in \emph{Workshop Machine Learning for Audio
  Signal Processing at {NIPS} ({ML}4Audio@{NIPS}17)}.

\bibitem{luo2018tasnet}
Y.~Luo and N.~Mesgarani, ``Tasnet: Time-domain audio separation network for
  real-time single-channel speech separation,'' in \emph{Acoustics, Speech and
  Signal Processing (ICASSP), 2014 IEEE International Conference on}.\hskip 1em
  plus 0.5em minus 0.4em\relax IEEE, 2018.

\bibitem{pascual2017segan}
S.~Pascual, A.~Bonafonte, and J.~Serra, ``Segan: Speech enhancement generative
  adversarial network,'' \emph{arXiv preprint arXiv:1703.09452}, 2017.

\bibitem{Fu2017Raw}
S.~W. Fu, Y.~Tsao, X.~Lu, and H.~Kawai, ``Raw waveform-based speech enhancement
  by fully convolutional networks,'' in \emph{2017 Asia-Pacific Signal and
  Information Processing Association Annual Summit and Conference (APSIPA
  ASC)}, Dec 2017.

\bibitem{rethage2018wavenet}
D.~Rethage, J.~Pons, and X.~Serra, ``A wavenet for speech denoising,'' 2018.

\bibitem{stoller2018wave}
D.~Stoller, S.~Ewert, and S.~Dixon, ``Wave-u-net: A multi-scale neural network
  for end-to-end audio source separation,'' \emph{arXiv preprint
  arXiv:1806.03185}, 2018.

\bibitem{fu2017end}
S.-W. Fu, Y.~Tsao, X.~Lu, and H.~Kawai, ``End-to-end waveform utterance
  enhancement for direct evaluation metrics optimization by fully convolutional
  neural networks,'' \emph{IEEE Transactions on Audio, Speech, and Language
  Processing}, 2018.

\bibitem{oord2016wavenet}
A.~v.~d. Oord, S.~Dieleman, H.~Zen, K.~Simonyan, O.~Vinyals, A.~Graves,
  N.~Kalchbrenner, A.~Senior, and K.~Kavukcuoglu, ``Wavenet: A generative model
  for raw audio,'' \emph{arXiv preprint arXiv:1609.03499}, 2016.

\bibitem{ronneberger2015u}
O.~Ronneberger, P.~Fischer, and T.~Brox, ``U-net: Convolutional networks for
  biomedical image segmentation,'' in \emph{International Conference on Medical
  image computing and computer-assisted intervention}.\hskip 1em plus 0.5em
  minus 0.4em\relax Springer, 2015, pp. 234--241.

\bibitem{wang2018end}
Z.-Q. Wang, J.~L. Roux, D.~Wang, and J.~R. Hershey, ``End-to-end speech
  separation with unfolded iterative phase reconstruction,'' \emph{arXiv
  preprint arXiv:1804.10204}, 2018.

\bibitem{taal2010short}
C.~H. Taal, R.~C. Hendriks, R.~Heusdens, and J.~Jensen, ``A short-time
  objective intelligibility measure for time-frequency weighted noisy speech,''
  in \emph{Acoustics Speech and Signal Processing (ICASSP), 2010 IEEE
  International Conference on}.\hskip 1em plus 0.5em minus 0.4em\relax IEEE,
  2010, pp. 4214--4217.

\bibitem{fu2018end}
S.-W. Fu, T.-W. Wang, Y.~Tsao, X.~Lu, and H.~Kawai, ``End-to-end waveform
  utterance enhancement for direct evaluation metrics optimization by fully
  convolutional neural networks,'' \emph{IEEE/ACM Transactions on Audio, Speech
  and Language Processing (TASLP)}, vol.~26, no.~9, pp. 1570--1584, 2018.

\bibitem{Kolbaek2018ontheequivalence}
M.~Kolbæk, Z.-H. Tan, and J.~Jensen, ``On the equivalence between objective
  intelligibility and mean-squared error for deep neural network based speech
  enhancement,'' \emph{arXiv preprint arXiv:1806.08404}, 2018.

\bibitem{fevotte2005bss_eval}
C.~F{\'e}votte, R.~Gribonval, and E.~Vincent, ``Bss\_eval toolbox user
  guide--revision 2.0,'' 2005.

\bibitem{Venkataramani2018Performance}
S.~Venkataramani, R.~Higa, and P.~Smaragdis, ``Performance based cost functions
  for end-to-end speech separation,'' in \emph{2018 Asia-Pacific Signal and
  Information Processing Association Annual Summit and Conference (APSIPA
  ASC)}, Dec 2018.

\bibitem{he2016deep}
K.~He, X.~Zhang, S.~Ren, and J.~Sun, ``Deep residual learning for image
  recognition,'' in \emph{Proceedings of the IEEE conference on computer vision
  and pattern recognition}, 2016, pp. 770--778.

\bibitem{hummersone2014ideal}
C.~Hummersone, T.~Stokes, and T.~Brookes, ``On the ideal ratio mask as the goal
  of computational auditory scene analysis,'' in \emph{Blind source
  separation}.\hskip 1em plus 0.5em minus 0.4em\relax Springer, 2014, pp.
  349--368.

\bibitem{huang2015joint}
P.-S. Huang, M.~Kim, M.~Hasegawa-Johnson, and P.~Smaragdis, ``Joint
  optimization of masks and deep recurrent neural networks for monaural source
  separation,'' \emph{IEEE/ACM Transactions on Audio, Speech, and Language
  Processing}, vol.~23, no.~12, pp. 2136--2147, 2015.

\bibitem{wang2018alternative}
Z.-Q. Wang, J.~Le~Roux, and J.~R. Hershey, ``Alternative objective functions
  for deep clustering,'' in \emph{2018 IEEE International Conference on
  Acoustics, Speech and Signal Processing (ICASSP)}.\hskip 1em plus 0.5em minus
  0.4em\relax IEEE, 2018.

\bibitem{smaragdis2017neural}
P.~Smaragdis and S.~Venkataramani, ``A neural network alternative to
  non-negative audio models,'' in \emph{Acoustics, Speech and Signal Processing
  (ICASSP), 2017 IEEE International Conference on}.\hskip 1em plus 0.5em minus
  0.4em\relax IEEE, 2017, pp. 86--90.

\bibitem{nugraha2016multichannel}
A.~A. Nugraha, A.~Liutkus, and E.~Vincent, ``Multichannel audio source
  separation with deep neural networks,'' \emph{IEEE/ACM Transactions on Audio,
  Speech, and Language Processing}, vol.~24, no.~9, pp. 1652--1664, Sept 2016.

\bibitem{wsj}
``Wall street journal 0 (wsj0) database,''
  \url{https://catalog.ldc.upenn.edu/ldc93s6a}.

\bibitem{allen1977unified}
J.~B. Allen and L.~R. Rabiner, ``A unified approach to short-time fourier
  analysis and synthesis,'' \emph{Proceedings of the IEEE}, vol.~65, no.~11,
  pp. 1558--1564, 1977.

\bibitem{cartwright_fast_2016}
M.~Cartwright, B.~Pardo, G.~J. Mysore, and M.~Hoffman, ``Fast and easy
  crowdsourced perceptual audio evaluation,'' in \emph{2016 IEEE International
  Conference on Acoustics, Speech and Signal Processing (ICASSP)}, March 2016,
  pp. 619--623.

\end{thebibliography}


%\newpage



%\vfill


\end{document}